\documentclass[iop, tighten, numberedappendix, letterpaper]{emulateapj}

\usepackage{verbatim}

\def\h2{$\rm H_2$}

\newcommand{\halpha}{H$\alpha$}
\newcommand{\hii}{H{\sc II}}
\newcommand{\hi}{H{\sc I}}

\newcommand{\hoi}{Ho~{\sc I}}
\newcommand{\hoii}{Ho~{\sc II}}

\begin{document}

\shortauthors{Weisz et al.}
\title{How Typical Are The Local Group Dwarf Galaxies?\footnotemark[*]}\footnotetext[*]{Based on observations made with the NASA/ESA Hubble Space Telescope, obtained from the Data Archive at the Space Telescope Science Institute, which is operated by the Association of Universities for Research in Astronomy, Inc., under NASA contract NAS 5-26555.}

\author{Daniel R. Weisz\altaffilmark{1,2},
Andrew E. Dolphin\altaffilmark{3},
Julianne J. Dalcanton\altaffilmark{1},
Evan D. Skillman\altaffilmark{2},
Jon Holtzman\altaffilmark{4},
Benjamin F. Williams\altaffilmark{1}, 
Karoline M. Gilbert\altaffilmark{1,9},
Anil C. Seth\altaffilmark{5},  
Andrew Cole\altaffilmark{6},
Stephanie M. Gogarten\altaffilmark{1},  
Keith Rosema\altaffilmark{1},
Igor D. Karachentsev\altaffilmark{7},
Kristen B.~W.~McQuinn\altaffilmark{2},
Dennis Zaritsky\altaffilmark{8}
}

\altaffiltext{1}{University of Washington; dweisz@astro.washington.edu}
\altaffiltext{2}{University of Minnesota}
\altaffiltext{3}{Raytheon}
\altaffiltext{4}{New Mexico State University}
\altaffiltext{5}{CfA Fellow, Harvard-Smithsonian Center for Astrophysics}
\altaffiltext{6}{University of Tasmania}
\altaffiltext{7}{Russian Academy of Sciences}
\altaffiltext{8}{Steward Observatory, University of Arizona}
\altaffiltext{9}{Hubble Fellow}

\begin{abstract}

We compare the cumulative star formation histories (SFHs) of Local Group (LG) dwarf galaxies with those in the volume-limited ACS Nearby Galaxy Survey Treasury (ANGST) sample (D $\lesssim$ 4 Mpc), in order to understand how typical the LG dwarf galaxies are relative to those in the nearby universe.  The SFHs were derived in a uniform manner from high quality optical color-magnitude diagrams constructed from Hubble Space Telescope imaging.  We find that the {\it mean} cumulative SFHs of the LG dwarfs are comparable to the mean cumulative SFHs of the ANGST sample for the three different morphological types (dwarf spheroidals/ellipticals: dSph/dE;  dwarf irregulars: dI; transition dwarfs: dTrans). We also discuss effects such as population gradients and systematic uncertainties in the stellar models that may influence the derived SFHs. Both the ANGST and Local Group dwarf galaxies show a consistent and strong morphology-density relationship, emphasizing the importance of environment in the evolution of dwarf galaxies.   Specifically, we confirm that dIs are found at lower densities and higher luminosities than dSphs, within this large sample. We also find that dTrans are located in similar environments to those occupied by dwarf irregular galaxies, but have systematically lower luminosities that are more comparable to those of dwarf spheroidals. The similarity of the SFHs and morphology-density relationships of the LG and ANGST dwarf galaxies suggests that the LG dwarfs are a good representation of dwarf galaxies in the local universe.

\end{abstract}


\section{Introduction}

Dwarf galaxies in the Local Group (LG) are among the most well-studied galaxies in the universe.  Detailed determinations of their kinematics, metallicities, and stellar contents serve as a basis for much of what we understand about the formation and evolution of both individual and groups of galaxies \citep[see reviews by][]{mat98, van00, tol09}.  In particular, we can directly determine the history of star formation and chemical evolution for individual galaxies  using Hubble Space Telescope (HST) observations of resolved stellar populations in nearby and LG galaxies  \citep[e.g.,][]{tol09}.

Although we often draw on results from LG studies to explain the evolution of galaxies in the broader universe, whether LG dwarf galaxies are representative of all dwarf galaxies remains an open question \citep[e.g.,][]{van00}.  The LG is a relatively dense environment with a specific history of mass accretion and interaction, which may have influenced dwarf galaxy evolution differently than nearby field or galaxy groups (e.g., the M81 Group).  Because we often extrapolate results from the LG to the more distant universe, it is important to establish the degree to which the LG dwarfs represent the broader dwarf galaxy population.

The Local Volume contains a diverse set of galaxies and environments, some of which have no analogs in the LG.  For example, the M81~Group is known to have undergone a recent major interaction \citep[e.g.,][]{yun94,yun99}, which has likely influenced star formation and gas-loss in the M81 Group dwarf galaxies  \citep[e.g.,][]{wei08, wal11}, and which may be responsible for the creation of new tidal dwarf galaxies \citep[e.g.,][]{mak02}.  At the opposite extreme in density, isolated `field' dwarf galaxies may have fewer or no interactions with more massive companions, which could result in distinctly different patterns of star formation and chemical evolution when compared to typical group members \citep[e.g.,][]{col07}.

\begin{deluxetable*}{lcccccc}
\tablecolumns{9}
\tabletypesize{\scriptsize}
\tablewidth{0pt}
\tablecaption{Local Group Dwarf Galaxy Sample}

\tablehead{   
\colhead{Galaxy} &   
    \colhead{Main} &
    \colhead{$M_{B}$} &
       \colhead{D} &
       \colhead{$A_{V}$} &
        \colhead{Type} &
      \colhead{$\Theta$}  \\
        \colhead{Name} &
    \colhead{Disturber} &
        \colhead{} &
       \colhead{(Mpc)} &
      \colhead{} &
      \colhead{} &
      \colhead{}  \\
           \colhead{(1)} &
    \colhead{(2)} &
    \colhead{(3)} &
        \colhead{(4)} &
       \colhead{(5)} &
      \colhead{(6)} &
      \colhead{(7)}   \\
}    

\tablecaption{Local Group Dwarf Galaxy Sample}
\startdata        
UrsaMin & MW & -7.13 &  0.08 & 0.11  & dSph & 3.3\\
LGS3 & M31 & -7.96 & 0.61 & 0.14 & dTrans & 1.7\\
And {\sc V} & M31 & -8.41 & 0.78 &  0.41 & dSph & 2.8\\
Draco & MW & -8.74 & 0.09 & 0.09 & dSph & 3.0 \\
Carina & MW & -8.97 & 0.10 & 0.21 & dSph & 2.7\\
Leo {\sc II} & MW &  -9.23 & 0.20 & 0.06 & dSph & 1.7\\
And {\sc III} & M31 &  -9.30 & 0.72 & 0.19 & dSph & 3.5\\
And {\sc II}& M31 & -9.33 & 0.65 & 0.21 & dSph & 2.4\\
Antlia & M31 & -9.38 & 1.30 & 0.24 & dTrans & -0.1 \\
Sculptor & MW &  -9.77 & 0.08 & 0.06 & dSph & 2.8\\
Cetus & M31 & -10.18 & 0.77 & 0.10 & dSph & 0.5\\
Phoenix & MW &  -10.22 & 0.41 & 0.05 & dTrans & 0.8\\
And {\sc VI} & M31 & -10.80 & 0.83 & 0.21 & dSph & 1.7\\
And {\sc I} & M31 & -10.87 & 0.76 & 0.18 & dSph & 3.7\\
Leo {\sc I} & MW & -10.97 & 0.25 &  0.12 & dSph & 1.5\\
DDO210 & M31 & -11.09 & 0.94 &  0.17 & dTrans & 1.6\\
Pegasus & M31 & -11.47 & 0.95 & 0.22 & dI & 1.2\\
SagDIG &  MW &-11.49 & 1.11 & 0.40 & dI & -0.3\\
Fornax & MW & -11.50 & 0.14 & 0.07 &  dSph & 2.3\\
And{\sc VII} & M31 & -11.67 & 0.94 & 0.64 & dSph & 2.0\\
LeoA & MW & -11.70 & 0.79 & 0.07 & dI & 0.1\\
Sagittarius & MW & -12.80 & 0.03 & 0.40 & dSph & 4.0\\
Tucana & MW & -12.94 & 0.86 &  0.11 & dSph &-0.1\\
SexA  & MW & -13.71 & 1.30 & 0.14 & dI & -0.6 \\
SexB  & MW & -13.88 & 1.40 & 0.10 & dI & -0.7 \\
WLM & M31 &-13.95 & 0.93 & 0.12 & dI & 0.3\\
N185 & M31 & -14.76 & 0.61 & 0.61 & dSph/dE & 3.5\\
N147 & M31 &-14.79 & 0.72 & 0.58 &  dSph/dE & 3.0\\
IC1613 & M31 & -15.57 & 0.74 & 0.08 & dI & 0.9
\enddata
\tablecomments{Properties of the sample of LG dwarf galaxies -- (1) Galaxy Names.  The SFH data for LGS3, Cetus, IC1613, Leo~A, Tucana, and IC1613 are from the LCID program \citep{gal07}.  The rest of the SFH data has been derived from the CMDs presented in \citet{dol05} and \citet{hol06}; (2) Most gravitationally influential neighbor \citep{kar04}; (3) Absolute Blue Magnitude; (4) Distance from TRGB or horizontal branch \citep{dol05}; (5) Foreground extinction \citep{sch98}; (6) Morphological Type \citep{mat98}; (7) Tidal Index \citep{kar04}.}
\label{lgtab}
\end{deluxetable*}

Directly comparing the stellar contents of LG dwarf galaxies to those in the nearby universe requires uniform data sets of comparable quality and size.  Historically, studies of resolved stellar populations in nearby dwarf galaxies have focused on small samples or individual galaxies, and have employed a variety of analysis techniques, leading to larger systematic uncertainties when comparing different studies.  Recent projects described by \citet[][]{hol06} and \citet[][]{dal09} have resulted in two uniformly processed multi-color photometric databases of the resolved stellar populations of dwarf galaxies in the LG and Local Volume, which, for the first time, allow a direct unbiased study of dwarf galaxies in a large volume, spanning a wide range of environments.  

In this paper, we present a comparison of the cumulative SFHs and morphology--density relationships of dwarf galaxies in the LG and the Local Volume.  We have measured the SFHs from uniformly processed photometry, using the same SFH code and stellar evolution models, minimizing the effects of potential systematics for comparison of the relative SFHs.  For the LG, we use the best fit SFHs from updated analysis of the LG sample \citet{dol10}, initially presented in \citet{dol05}.  SFHs for the Local Volume are taken from the measured SFHs of ANGST dwarf galaxies \citep{wei10}. We specifically address the question of whether or not the most likely cumulative SFHs of the LG dwarfs are comparable to those of the ANGST sample.  In addition to comparing the \emph{mean} cumulative SFHs, we also consider effects that can potentially introduce biases in the measured SFHs, such as population gradients and systematic uncertainties in the underlying stellar models.   

This paper is organized as follows.  In \S \ref{data}, we summarize the sample selection and data. The technique of measuring SFHs from optical color-magnitude diagrams (CMDs) is briefly reviewed in \S \ref{sfhs}.  In \S \ref{discuss}, we address the question of whether the SFHs of LG dwarf galaxies are consistent with those in the Local Universe, by comparing cumulative SFHs from the ANGST and LG dwarf galaxy samples.  We then examine the morphology--density relationship for both LG and and ANGST dwarfs, and discuss demographic differences in the samples in \S \ref{morph}.  Cosmological parameters used in this paper assume a standard WMAP-7 cosmology \citep{jar10}.

\section{The Data}
\label{data}

\subsection{The Local Group Dwarf Galaxy Sample}

For this comparison, we consider a sample of LG dwarf galaxies based on those discussed in \citet{mat98} (see Table \ref{lgtab}).  All galaxies have multi-color optical imaging taken with either the Advanced Camera for Surveys \citep[ACS;][]{for98} or the Wide Field Planetary Camera 2 \citep[WFPC2;][]{hol95} aboard HST.  Following the convention of \cite{mat98}, we have excluded the LMC and SMC from this paper. Similarly, we have omitted NGC~3109, NGC~205, NGC~6822, and NGC~55 as these are sufficiently luminous galaxies that their status as dwarfs  is ambiguous \citep[e.g.,][]{hod71, ski96, mat98, wei10}. Similarly, we have excluded  `ultra-faint' LG dwarf galaxies \citep[e.g.,][]{bel06} and Andromeda companions discovered more recently than And~{\sc VII} \citep[e.g.,][]{zuc04} from this comparison; these sets of galaxies do not yet have publicly available HST photometry that has been processed in the same way as the galaxies we consider in this paper.  The resulting LG sample covers a range of $M_{B}$ from $-$7.13 (Ursa Minor) to $-$15.57 (IC1613).

We divide the sample into three categories, dwarf spheroidal/elliptical (dSph), dwarf irregular (dI), and transition dwarf (dTrans) galaxies according to the morphological classifications in \citet{mat98}. We combine dwarf spheroidals and the two dwarf ellipticals \citep[dE; NGC~147 and NGC~185; e.g.,][]{mat98} into a canonical gas-poor galaxy (dSph) category.   Following the convention of \citet{mat98}, we designate dTrans as those which have detectable amounts of \hi\ and no \emph{significant} recent star formation, as measured by \halpha. We note that \citet{mat98} classifies Pegasus as a dTrans, but \citet{ski03a} detect significant levels of \halpha, and thus we consider it to be a dI.  We also exclude IC~10 due to its low galactic latitude and high foreground extinction \citep[$A_V$ $\sim$ 2.5;][]{sch98}, making both its distance and SFH highly uncertain.  In total, the LG sample has 18 dSphs, 7 dIs, and 4 dTrans (Table \ref{lgtab}).

\begin{deluxetable*}{lcccccc}
\tablecolumns{7}
\tablewidth{0pt}
\tablecaption{ANGST Dwarf Galaxy Sample}

\tablehead{   
\colhead{Galaxy} &   
    \colhead{Main} &
    \colhead{$M_{B}$} &
       \colhead{D} &
       \colhead{$A_{V}$} &
        \colhead{Type} &
      \colhead{$\Theta$} \\
        \colhead{Name} &
    \colhead{Disturber} &
        \colhead{} &
       \colhead{(Mpc)} &
      \colhead{} &
      \colhead{} &
      \colhead{}  \\
           \colhead{(1)} &
    \colhead{(2)} &
    \colhead{(3)} &
        \colhead{(4)} &
       \colhead{(5)} &
      \colhead{(6)} &
      \colhead{(7)}   \\
}    

\startdata

KK230 &  M31 & -8.49 & 1.3 & 0.04 &  dTrans & -1.0 \\
KKR25 &  M31 & -9.94 & 1.9 & 0.03 & dTrans & -0.7 \\
FM1 & M82 & -10.16 & 3.4 & 0.24 & dSph/dE & 1.8   \\
KKH86 &  M31 & -10.19 & 2.6 & 0.08 & dI & -1.5 \\ 
KKH98 &  M31 &-10.29 & 2.5 & 0.39 & dTrans & -0.7 \\
BK5N  & N3077 & -10.37 & 3.8  & 0.20 & dSph/dE & 2.4 \\
Sc22  & N253 & -10.39 & 4.2 & 0.05 & dSph & 0.9 \\
KDG73  &M81 & -10.75 & 3.7 & 0.06 & dTrans & 1.3 \\
IKN  & M81 & -10.84 & 3.7 & 0.18 & dSph & 2.7 \\
E294-010  & N55 & -10.86 & 1.9 & 0.02 & dTrans & 1.0 \\
E540-032  & N253 &-11.22 & 3.4 & 0.06 & dTrans & 0.6 \\
KKH37  & I342 & -11.26 & 3.4  & 0.23 &dI & -0.3 \\
KDG2  & N253 & -11.29 & 3.4 & 0.07 & dTrans & 0.4 \\
UA292 & N4214 & -11.36 & 3.1 & 0.05 & dI & -0.4 \\
KDG52  & M81 & -11.37 & 3.5 & 0.06 & dTrans & 0.7 \\
KK77  & M81 & -11.42 & 3.5 & 0.44 & dSph/dE & 2.0 \\
E410-005  & N55 & -11.49 & 1.9 & 0.04 & dTrans & 0.4 \\
HS117 &  M81 & -11.51 & 4.0 & 0.36 & dI & 1.9 \\
DDO113 & N4214 & -11.61 & 2.9 & 0.06 &dI & 1.6 \\
KDG63  & M81 & -11.71 & 3.5 & 0.30 & dSph/dE & 1.8 \\
DDO44  & N2403 &  -11.89 & 3.2 & 0.13 & dSph & 1.7 \\
GR8  &M31 & -12.00 & 2.1 & 0.08 & dI & -1.2 \\
E269-37   & N4945 & -12.02 & 3.5 & 0.44 & dSph & 1.6 \\
DDO78  & M81 & -12.04 & 3.7 &0.07 & dSph & 1.8 \\
F8D1 & M81 & -12.20 & 3.8 & 0.33 & dSph/dE & 3.8 \\
U8833  & N4736 &-12.31 & 3.1 & 0.04 & dI & -1.4 \\
E321-014  & N5128 & -12.31 & 3.2 & 0.29 & dI & -0.3 \\
KDG64  & M81 & -12.32 & 3.7 & 0.17 & dSph/dE & 2.5 \\
DDO6  & N253 & -12.40 & 3.3 & 0.05 & dTrans & 0.5 \\
DDO187  & M31 & -12.43 & 2.3 & 0.07 & dI & -1.3 \\
KDG61 & M81 &-12.54 & 3.6 & 0.23 & dSph/dE & 3.9 \\
U4483  & M81 & -12.58 & 3.2 & 0.11 & dI & 0.5 \\
UA438  & N55 & -12.85 & 2.2 & 0.05 & dTrans & -0.7 \\
DDO181  & M81 &  -12.94 & 3.0 & 0.02 & dI & -1.3 \\
U8508  & M81 & -12.95 & 2.6 & 0.05 & dI & -1.0 \\
N3741  & M81 & -13.01 & 3.0 & 0.07 & dI & -0.8 \\
DDO183  & N4736 & -13.08 & 3.2 & 0.05 & dI & -0.8 \\
DDO53  & M81 & -13.23 & 3.5 & 0.12 & dI & 0.7 \\
DDO99  & N4214 & -13.37 & 2.6 & 0.08 & dI & -0.5 \\
N4163  & N4190 & -13.76 & 3.0 & 0.06 & dI & 0.1 \\
DDO125  &  N4214 & -14.04 & 2.5 & 0.06 & dI & -0.9 \\
E325-11 &   N5128 & -14.05 & 3.4 & 0.29 &dI & 1.1  \\
DDO190  & M81 & -14.14 & 2.8 & 0.04 & dI & -1.3 \\
\hoi  & M81 &  -14.26 & 3.8 & 0.15 & dI & 1.5 \\
DDO165  & N4236 & -15.09 & 4.6 & 0.08  &dI & 0.0 \\
IC5152 & M31 &-15.67 & 1.9 &  0.08 &  dI & -1.1\\
N2366  & N2403 &  -15.85 & 3.2 & 0.11 & dI & 1.0 \\
\hoii  & M81 & -16.57 & 3.4 & 0.10 & dI & 0.6 
\enddata
\tablecomments{\small{Properties of the sample of ANGST dwarf galaxies -- (1) Galaxy Names; (2) Most gravitationally influential neighbor \citep{kar04}; (3) Absolute Blue Magnitude; (4) Distance from TRGB \citep{dal09}; (5) Foreground extinction \citep{sch98}; (6) Morphological Type \citep{kar04, wei10}; (7) Tidal Index \citep{kar04}.}}
\label{angsttab}
\end{deluxetable*}

The SFHs used for this comparison are based on CMDs presented in \citet{dol05} and \citet{hol06} that have been re-measured and analyzed \citep{dol10} with updated Padova stellar evolution models \citep{mar08}.  These SFHs are based on photometry of HST/WFPC2 imaging that was uniformly processed using HSTPHOT \citep{dol00} as part of the Local Group Stellar Populations Archive\footnote[2]{http://astronomy.nmsu.edu/holtz/archival/html/lg.html} \citep{hol06}.  

Five of the LG dwarfs (Cetus, Tucana, IC~1613, Leo~A, and LGS3) were more recently observed with ACS as part of the Local Cosmology with Isolated Dwarfs program \citep[LCID;][]{gal07}, and have significantly deeper CMDs compared to the corresponding WFPC2 photometry \citep[e.g.,][]{col07, mon10a, mon10b}.  Photometry of ACS observations was performed with DOLPHOT\footnote[3]{http://purcell.as.arizona.edu/dolphot/}, an update of HSTPHOT with an ACS specific module, which allows us to include these five SFHs without compromising uniformity.

\subsection{The ANGST Dwarf Galaxy Sample}

The ACS Nearby Galaxy Survey Treasury \citep[ANGST;][]{dal09}\footnote[4]{http://archive.stsci.edu/prepds/angst/}  sample contains dwarf galaxies located beyond the zero velocity surface of the LG \citep{van00} and within $D$ $\sim$ 4 Mpc (see Table \ref{angsttab}).  The sample contains a mixture of field and group galaxies, the latter of which are located in the M81~Group (D$_{M81}$ $\sim$ 3.6 Mpc) and in the direction of the NGC~253 clump (D$_{N253}$ $\sim$ 3.9 Mpc) in the Sculptor Filament \citep{kar03}.  For comparison, we have selected all dSphs, dIs, and dTrans for the ANGST dwarf sample, as defined in \citet{wei10}.  We note that several ANGST galaxies (Antlia, GR8, Sex~A, Sex~B, IC~5152, and UGCA~438) are considered to be in the LG of \citet{mat98}.  Of these galaxies, Antlia, Sex~A, and Sex~B are on the periphery of the LG, and we classify them as part of the LG sample for the purposes of this study. TRGB distance determinations \citep[e.g.,][]{dal09} place IC~5152, GR8, and UA~438 clearly beyond the zero velocity surface of the LG, and we thus include them in the ANGST sample for this comparison.

The ANGST sample contains several reported dEs, all within the M81~Group: F8D1, BK5N, KDG~61, KDG~64, KK~77, FM~1, and DDO~71 \citep{cal98, DaC07}.  We include these in the dSph category.  Although the term dE has often been used to describe early type dwarf galaxies outside the LG \citep[e.g.,][]{geh06}, here we adopt the term dSph, which has historically been used in LG studies \citep[e.g.,][and references therein]{mat98} as a number of the early type ANGST dwarf galaxies have luminosities and CMDs that are similar to dSphs in the LG.  We discuss potential differences between dEs and dSphs in this sample in \S \ref{morph}.

The faint end of the ANGST galaxy luminosity function is likely not complete due to selection biases against identifying faint low luminosity surface brightness galaxies.  For example, the ANGST sample does not include a number of recently discovered low surface brightness M81~Group galaxies \citep{chi09}, as they do not yet have HST imaging publicly available. The final sample of ANGST galaxies in this paper includes 12 dSphs, 25 dIs, and 11 dTrans, spanning a range in $M_{B}$ from $-$8.49 (KK230) to $-$16.57 (\hoii). 

The SFHs of the ANGST dwarf galaxies are presented in \citet{wei10}.  Data reduction and analysis, i.e., photometry and SFHs, have been done in a manner consistent with the LG sample, making it possible for a direct comparison.

\section{Measuring the Star Formation Histories}
\label{sfhs}

We briefly summarize the technique of measuring a SFH from a CMD, based on more detailed discussions of the methodology described in \citet{dol02}, \citet{dol05}, and \citet{wei10}.   For this technique, we specify an IMF, binary fraction, and allowable ranges in age, metallicity, distance, and extinction.  Photometric errors and completeness are characterized by artificial star tests.  From these inputs, the code generates many synthetic CMDs spanning the desired age and metallicity range.  For this work, we used synthetic CMDs sampling stars with logarithmic age and metallicity spreads of 0.05 and 0.1 dex, respectively.  These individual synthetic CMDs were then linearly combined along with a model foreground CMD to produce a composite synthetic CMD.  The linear weights on the individual CMDs are adjusted to obtain the best fit as measured by a Poisson maximum likelihood statistic; the weights corresponding to the best fit are the most probable SFH.  This process can be repeated at a variety of distance and extinction values to solve for these parameters as well. 

In this study, we consider the cumulative SFHs, which give the fraction of stellar mass formed prior to a given time.  We prefer this presentation to the more traditional plot of star formation rate (SFR) vs. time, as the cumulative SFHs allow galaxies of different masses to be directly compared, and cumulative SFH measurements are not subject to covariant SFRs in adjacent time bins (see Appendix A of \citealt{wei10}).  We specifically compute an average cumulative SFHs per morphological type for both the LG and ANGST samples, which is the unweighted mean of the individual cumulative SFHs of galaxies within each group.  This scheme weights all galaxies equally, so that the resulting SFH indicates what's `typical' for a particular morphological type.  The resulting mean cumulative SFHs are shown in Figure \ref{cum_sfh}.  

We assign uncertainties to the mean cumulative SFHs using the method of error analysis  outlined in \citet{dol02} and \citet{wei10}.  First, we divide the total uncertainty in the derived SFH into random and systematic components.  Random uncertainties are due to Poisson uncertainties of the number of stars used to derive the SFR in a given time bin.  The random uncertainties include observational effects, such as photometric errors and completeness corrections, which can affect the number of stars in each model time bin.  To compute the random component of the uncertainty in the mean cumulative SFH, we add the uncertainties in the SFRs per time bin from the individual SFHs in quadrature.  These random uncertainties are shown as the error bars in the left panel of Figure \ref{cum_sfh}.  For well-populated CMDs, including most of those used to measure the SFHs considered in this paper, the amplitude of random uncertainties are typically $\lesssim$ 10\%.

The second source of uncertainty we consider are systematic uncertainties.  Systematics are introduced into the measured SFHs through biases in the underlying stellar models that change as a function of photometric depth of the observed CMD.  The photometric depth of a CMD determines the presence of age sensitive features (e.g., ancient MS turn-off, horizontal branch) available for measuring a SFH \citep[e.g.,][]{apa09, wei10}.  For extremely deep CMDs that include the ancient main sequence turn off (MSTO),  systematics are typically small, because of the reliability of MS stellar evolution models.  For shallower CMDs, including the majority of those used for SFH measurements in this paper, we must rely on the evolved star populations (e.g., RGB, HB) to constrain the SFH.  The physics of evolved stars is not yet fully understood, and consequently SFHs derived from CMDs that primarily contain evolved stars may be systematically biased \citep[e.g.,][]{gza05}.  Appendix B in \citet{wei10} demonstrates the effect of systematic biases on SFH measurements.

For comparison of SFHs in this paper, we are interested in the differential systematic uncertainties between the LG and ANGST samples.  Specifically, we wish to understand how the different typical photometric depths between the two samples affects the derived SFHs.  To estimate the systematic uncertainties, we follow the procedure outlined Appendices B and C in \citet{wei10}.

Briefly, this technique uses the differences in SFHs derived using the BaSTI \citep{pie04} and Padua stellar evolution models as an estimate of systematic errors.  In the case of the LG and ANGST galaxies, we did the following exercise to gauge the systematic uncertainties in the derived SFHs.  First, we simulated a CMD assuming constant SFH using the BaSTI stellar evolution models at photometric depths equivalent to those in the LG and ANGST samples.  We then recovered the SFH of each simulated CMD using the Padua models.  Next, we took the cumulative SFH of each realization and computed the mean of the ensemble population.  For example, for each of the 18 ANGST dSphs, we simulated a CMD assuming a constant SFH to the depth of the observed CMD, using the BaSTI models, and then recovered the SFH using the Padua models.  From these 18 realizations, we then computed the mean cumulative SFH for this collection of simulations.  This exercise was then repeated for all the other morphological sub-groups, e.g., LG dSphs, ANGST dIs, etc.

We now want to compare the differences in the observed cumulative SFHs with the size of the systematic uncertainties.  To do this, we simply take the difference of the observed cumulative SFHs and the simulated cumulative SFHs, and compare the amplitude of the difference per time bin.  As a concrete example, we can consider the systematics associated with the dSphs.  We first compute the difference in the mean cumulative SFHs, and show them as the black points in Panel (a) of Figure \ref{cum_sfh}.  The error bars on each point are the random uncertainties.  We then compute the difference in the mean recovered SFHs of the LG and ANGST dSphs samples.  This difference is plotted as the red points in Panel (a) of Figure \ref{cum_sfh}.  In general, if the value of the red points are greater than those of the black points, the systematic uncertainties are larger than the difference in the measured SFHs.  Note that this exercise is an equivalent to a single Monte Carlo realization for each ensemble, and thus provides an \emph{estimate} of the systematic uncertainties. 

We consider an additional source of uncertainty, which is due to the limited areas ample by the LG HST observations.  From studies of LG dwarfs, it has long been known that the central regions of dwarfs typically have larger numbers of young or intermediate age stars, while populations further from the center tend to have older average ages \citep[e.g.,][]{hod73, hun86, irw95, mat98, hid09}.  Because of their close proximity, LG dwarf galaxies have large angular areas relative to HST's field of view and,  in most cases, the deepest available HST observations target the central regions of dwarf galaxies \citep[e.g.,][]{hol06}.  The implications for this study are that  the LG SFHs derived for the central regions may not be representative of the entire galaxies.  In contrast, the ANGST galaxies are distant enough that a single HST field typically covers the entire extent of a dwarf galaxy \citep{wei10}.  The net effect of this bias is that SFHs of the LG dwarfs are likely weighted toward somewhat younger ages.  However, quantifying such an effect on SFHs must wait for wide-field CMDs that reach the ancient MSTO.

Finally,  as evidenced by the amplitude of the systematic uncertainties (shown on the right side of Figure \ref{cum_sfh}), we cannot reliably compare SFHs prior to z $\sim$ 1 ($\gtrsim$ 7.6 Gyr ago). At these early epochs, only ultra-deep CMDs reaching the ancient main sequence turnoff can reliably decipher the details of the ancient SFHs, e.g., the epoch of the peak ancient SFR.  Although some such CMDs are available within the LG sample, none are present for the more distance galaxies in the ANGST sample.  Thus, while the two samples are generally comparable in their cumulative SFHs, studies of the detailed ancient SFHs indicate that galaxies can show wide variation in their patterns of star formation at the oldest epochs \citep[e.g.,][]{col07, mon10a, mon10b}.  These early variations may show correlations with environment or other galaxy properties.  However, we are not able to discern such differences for the vast majority of galaxies in our samples.

\section{Discussion}
\label{discuss}

\begin{figure*}[t]
\begin{center}
\epsscale{1.1}
\plottwo{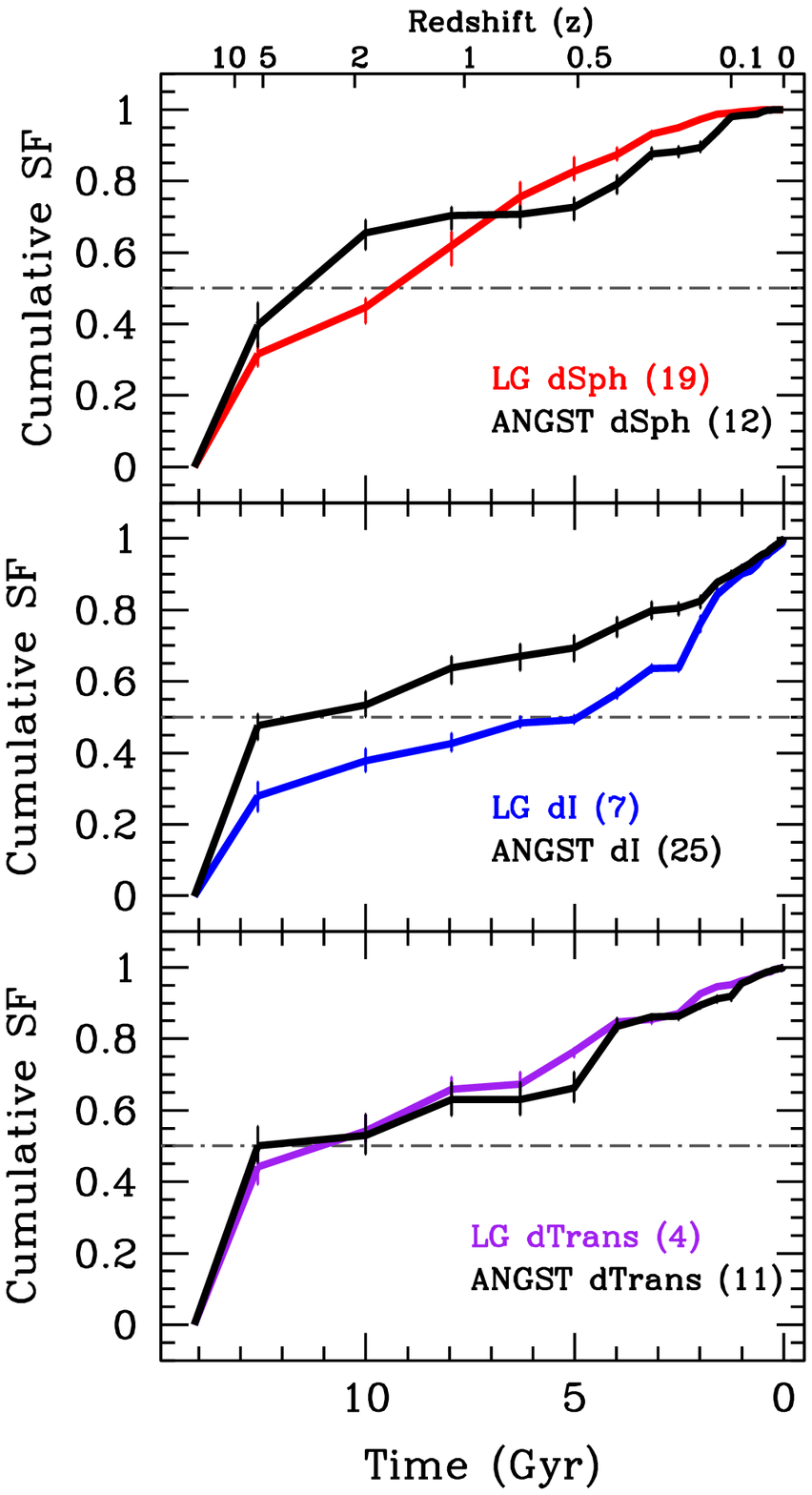}{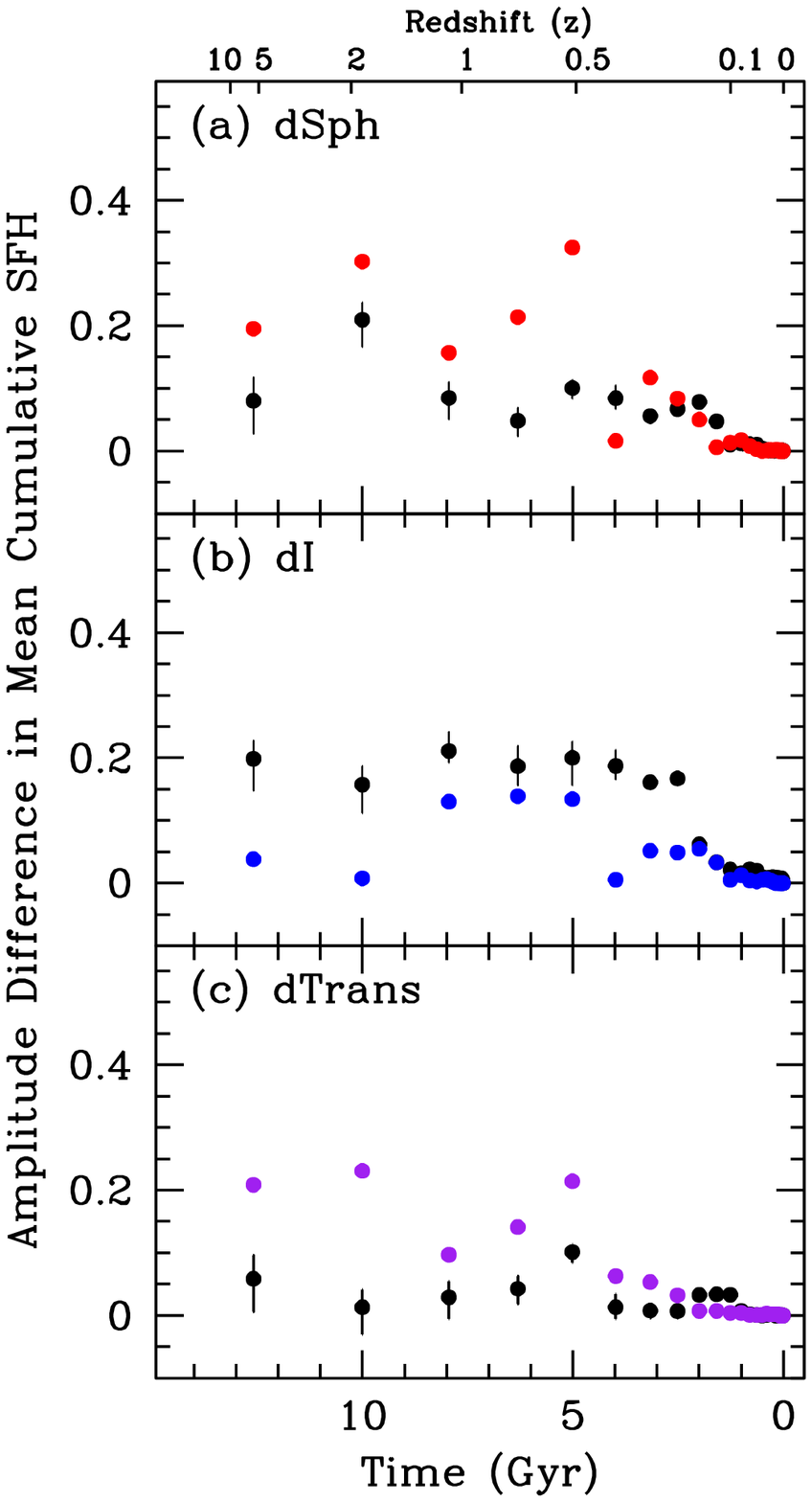}
\caption{Left -- The cumulative SFHs of the LG and ANGST sample dwarfs galaxies with random uncertainties shown as error bars.  Right -- The absolute value of the difference between the mean cumulative SFHs of the ANGST and LG samples, per morphological type, with random uncertainties (black).  The estimated systematic effects in each sample are plotted as colored points.  These values were determined by considering the absolute value of the difference between the simulated input SFH using BASTI models and the recovered SFH using Padova models.  The number and photometric depths of recovered SFHs were chosen to match to the appropriate sample (e.g., 18 and 12 SFHs with photometric depths of the LG and ANGST dSphs were used to measure the values of the red points).  Because the amplitude of the systematic uncertainties increase dramatically at old ages, points in the cumulative SFHs older than z $\sim$ 1 (7.6 Gyr ago) cannot be reliably compared. See \S \ref{sfhs} for a more detailed discussion. }
\label{cum_sfh}
\end{center}
\end{figure*}

In the left panel of Figure \ref{cum_sfh}, we compare the mean cumulative SFHs of the ANGST and LG samples per morphological type.  The plotted error bars represent only the random uncertainties in the SFHs, i.e., .  The expected amplitude of systematic uncertainties are shown in the right panel.  

\subsection{Comparing the ANGST and LG Star Formation Histories}
\label{comp_sfhs}

We first compare the mean cumulative SFHs of the ANGST and LG samples.  Taken at face value, the SFHs in the left panel of Figure \ref{cum_sfh} indicate that the cumulative SFHs of dSphs and dTrans are generally comparable, for the ANGST and LG volumes.  In particular, the typical dSph or dTrans galaxy in either sample formed $\sim$ 70\% of its stars by z $\sim$ 1 (7.6 Gyr ago), the look back time to which the typical ANGST CMD is sensitive \citep{wei10}.  At more recent times, we see some differences between the cumulative SFHs of the two dSph populations.  The LG dSphs appear to have formed a slightly higher percentage of stars by z $\sim$ 0.2, but the two sample show agreement for times more recent than z $\sim$ 0.1.

In contrast, the cumulative SFHs of LG and ANGST dIs do not generally agree within the random uncertainties. In particular, the typical LG dI has formed $\sim$ 50\% of its stars by z $\sim$ 1 (7.6 Gyr ago), which the typical ANGST dI has formed $\sim$ 65\% by the same epoch, although we note that accounting for effects of population gradients would likely increase the percent of stellar mass formed an ancient times for the LG dIs, brining the samples into closer agreement.

For meaningful comparison of the SFHs, we cannot take the mean SFHs at face value, and instead must consider the influence of systematic effects on the derived SFHs.  As outlined \S \ref{sfhs} we have estimated the systematic effects as a function of varying
numbers of galaxies observed at varying depths of photometry, and find that, overall,
the differences between the LG and ANGST dwarf galaxy SFHs are typically smaller than 
the systematic and statistical uncertainties, suggesting that the SFHs of the two samples 
are likely not very different.    We see that the systematic effects on the mean cumulative SFHs are largest for
the the dSph galaxy samples (i.e., the colored points are usually larger than 
the black points) for ages older than 4 Gyr. For the dTrans samples the differences
are very comparable to the systematic uncertainty estimates, and for the dIs samples, 
the systematic uncertainties are generally less than 
the observed differences in the mean cumulative SFHs.  Thus, the SFHs of the dSphs and dTrans in the LG and ANGST volumes  appear to be be fairly similar.

\begin{figure*}[t]
\begin{center}
\epsscale{1.0}
\plotone{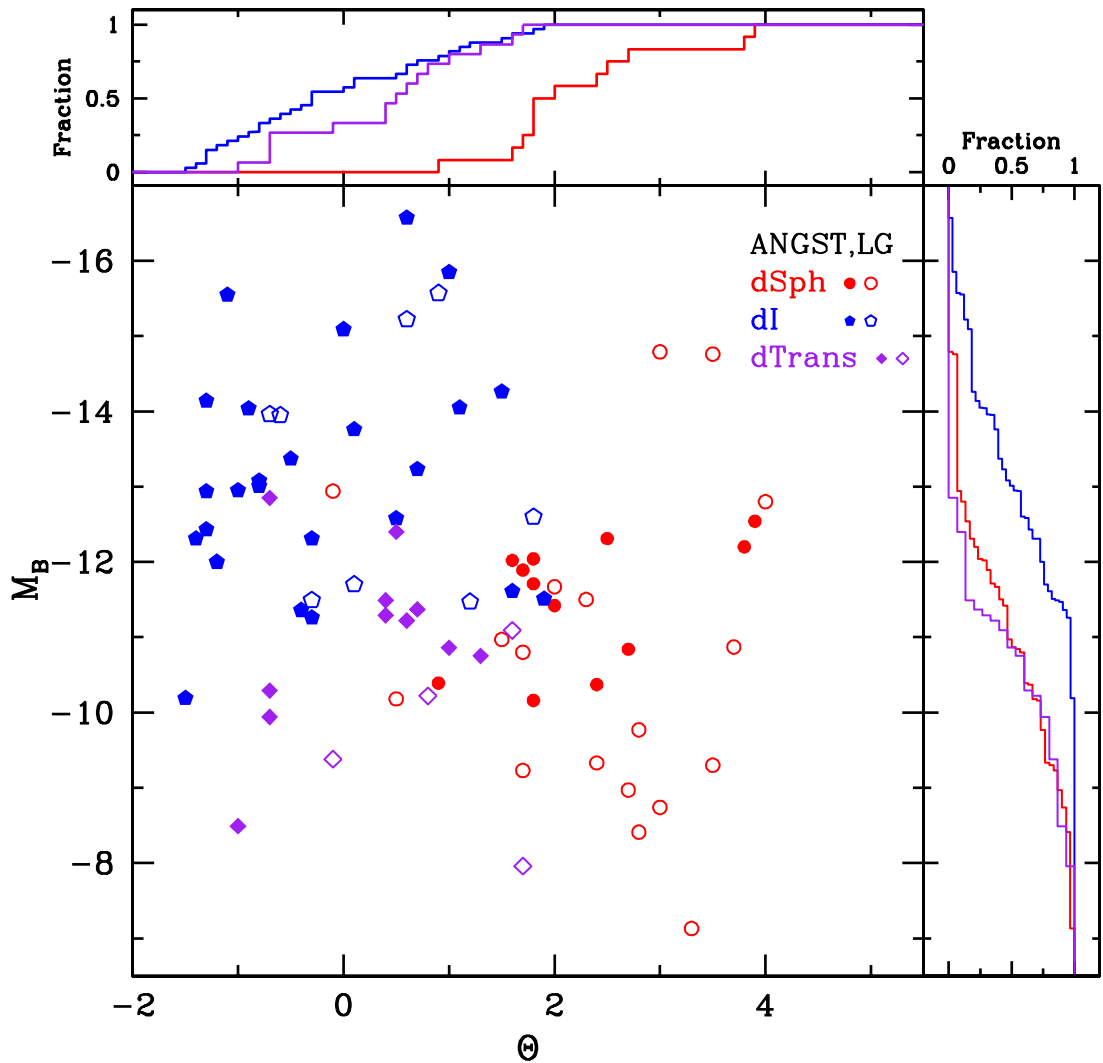}
\caption{The observed morphology--density relationship for the LG (open symbols) and ANGST (filled symbols) dwarf galaxies.  Negative values of $\Theta$ represent isolated galaxies, while positive values represent typical group members.  We see that both LG and ANGST dIs typically have negative tidal indices, while dSphs generally have positive tidal indices.  dTrans have intermediate values, namely that they have a similar luminosity distribution to dSphs, but are located in low density regions, similar to dIs (see \S \ref{morph}). The strong morphology--density relationship supports suggestions that environmental factors are important in the evolution of dwarf galaxies in both the Local Group and Local Volume.}  
\label{mdr}
\end{center}
\end{figure*}

Interpreting the differences in the cumulative SFHs of the LG and ANGST dIs is not straightforward.  On one hand, the differences their mean cumulative SFHs seem to exceed the estimates of the systematic uncertainties.  Taken at face value, this suggests that the dIs in the two samples are not consistent.  However, there are two effects which convolute this interpretation.  First, is the effect of population gradients, which tend to bias the ancient SFHs of LG dIs to somewhat younger ages.  Because we cannot quantify the size of this effect, we only note that the cumulative SFHs of the LG dIs are likely lower limits.  The second issue is interpreting the meaning of the systematic uncertainties.  We have used the differences in SFHs recovered from two sets of stellar models as a proxy for systematic uncertainties.  Yet, these models share a degree of similarity in their underlying physics, for example nuclear reaction rate and stellar atmospheres, which may make similarities in these models larger than the similarity between any one model and observed data.  Thus, in some sense, the systematics we consider in this study are also lower limits due to the inherent similarity of the underlying stellar models.  

While it is not clear if the observed difference in the SFHs of the LG and ANGST dIs is indicative of different modes of SF or simply the result of several biases, we do suggest that, at minimum,  these SFHs are not drastically different.

\subsection{Variations Among Morphological Types in the Combined Local Group and Local Volume Sample}
\label{morph}

The comparison shown in \S \ref{comp_sfhs} suggests that the differences in the SFH between the LG and ANGST dwarfs are not pronounced.  We therefore now consider the SFH and morphological properties of the combined sample of $\sim$ 80 galaxies.

From the SFH perspective, the SFHs of the entire LG and ANGST dwarf population confirm the findings of \citet{wei10}.  Namely, we verify that the dSphs, dTrans, and dIs have very similar cumulative SFHs until z $\sim$ 0.7.  Subsequent to this time, the average dSphs formed a higher fraction of its stellar mass, and appear to have little SF more recently than z $\sim$ 0.1.  In contrast, dIs formed higher fractions of their stellar mass at more recent times.  dTrans straddle the middle between the dSphs and dIs.

The combined LG-ANGST dwarf sample also gives us excellent dynamic rang for probing the morphology density relationship.  In Figure \ref{mdr}, we see that the LG and ANGST samples show clear and well-matched morphology--density relationships.  The dIs in both samples span a broad range of predominantly negative tidal indices \citep[$\Theta$, a useful probe of local density;][]{kar04} indicating that dIs are preferentially found away from dense environments.  This result is consistent with previous studies of both the LG \citep[e.g.,][]{mat98, van00} and dwarf galaxies in the wider universe \citep[e.g.,][]{ski03a, geh06}.  The ANGST sample has more dIs with $\Theta$ $<$ $-$1.5, which is due to the large field population of the sample.  dSphs in the two samples typically have positive tidal indices and fainter values of $M_{B}$ when compared to the dIs.  A Kolmogorov-Smirnov (KS) test confirms that dIs and dSphs are not drawn from the same distribution in either environment or luminosity, with a less than $4\times10^{-5}$ probability of having identical distributions.
  
In contrast with the sharp division in properties between dIs and dSphs, dTrans have properties that are intermediate between the two classes.  dTrans galaxies occupy environments that are statistically indistinguishable from those of dIs, according to a KS test.  Their environments are statistically distinct from those occupied by dSphs, with only a $2\times10^{-6}$ probability of being drawn from the same distribution of $\Theta$.  The luminosities of the dTrans galaxies have the opposite behavior. Their luminosities are statistically indistinguishable from those of dSphs, but are statistically lower luminosity than dIs, with only a $10^{-5}$ probability of being drawn from the same distribution of luminosities.  

To first order, the results above suggest that the majority of dTrans are consistent with being the low mass end of the dI population.  At these low masses, there are seldom more than a few \hii\ regions per galaxy \citep[e.g.,][]{mat98, ski03a} leading to a dTrans classification.  However, we note that gas-rich galaxies with little \halpha\ can form in multiple ways \citep[e.g.,][]{wei10} suggesting that not every dTrans has a similar evolutionary history.  Those with higher tidal indices may be interacting galaxies in the process of transforming to gas-poor dSphs \citep[e.g.,][]{gre03}. We refer the reader to \citet{wei10} for a more detailed discussion of the nature of dTrans. 

Although we have grouped all early type dSphs and dEs together into a single `dSph' category, it is interesting to search for distinguishing characteristics between the two morphological types. Figure \ref{mdr} reveals that the two dEs in the LG, ($M_{B}=-$14.79) and NGC~185 ($M_{B}=-$14.76), are conspicuously more luminous than other early type dwarfs.  Their intrinsic brightness is likely linked to the relatively large amount of intermediate age SF \citep[e.g.,][]{han97, mat98, but05, dol05}.  In contrast, the seven dEs in the ANGST volume \citep[F8D1, BK5N, KDG~61, KDG~64, KK~77, FM~1, and KDG~63;][]{cal98, DaC07}, span a broad range of lower luminosities from FM~1 ($M_{B}=-$10.16) to KDG~61 ($M_{B}=-$12.54).  

Examining the global SFHs of the ANGST dEs, we find that the two lowest luminosity ANGST dEs (FM~1 and BK5N) are dominated by stellar populations older than $\sim$ 10 Gyr, while the more luminous dEs (KK~77, KDG~63, F8D1, KDG~64, and KDG~61) have larger fractions of intermediate SF, albeit at a lower absolute SFR than the two LG dEs \citep[e.g.,][]{wei10, dol05}.  The predominance of old stellar populations in the two low luminosity dEs is more consistent with the LG and ANGST dSph populations, suggesting a dSph designation would be more accurate.

Interestingly, two of the most luminous ANGST dEs (F8D1 and KDG~61) have high tidal indices ($\Theta$ $\sim$ 4) that are comparable to the LG dEs.  This suggests that, like NGC~147 and NGC~185 which have a history of interaction with M31 (\citealt{mat98} and references therein), F8D1 and KDG~61 may have have experienced interactions with their most gravitationally influential neighbor, M81.  The two LG dEs may be examples of formerly gas-rich, rotationally supported dIs that are in the process of being transformed into gas-poor, pressure supported dSphs/dEs \citep[e.g.,][]{geh10}.  Further detailed study of the stellar kinematics of F8D1 and KDG~61 may help determine if these galaxies are undergoing a similar process, and may provide insight into the possible physical mechanisms \citep[e.g., ram pressure, stellar mass loss;][]{may01a, may01b, kaz10} driving this transformation.

\section{Summary}

We present a comparison of the cumulative SFHs of LG and ANGST dwarf galaxies from high quality CMDs based on HST imaging.  In order to minimize systematics, the SFHs from each samples were derived using identical techniques and input parameters.  The typical LG and ANGST dwarf galaxy appear to have formed the majority of their stellar mass prior to z $\sim$ 1 (7.6 Gyr ago).  Comparing the SFHs between the morphological types, we find that the dSphs and dTrans in the ANGST and LG samples have similar cumulative SFHs.  In contrast, the LG and ANGST dIs exhibit differences in their SFHs at ancient times, when taken at face value.  We show that the observed differences are likely lower limits, due to the effects of population gradients, and the lower limit systematic uncertainties we have calculated.  

Give that similarity of the LG and ANGST dwarf populations, we combine them into a single large sample.  This combined sample exhibits a well-defined morphology--density relationship, with dIs showing higher degrees of isolation than dSphs.  A KS test reveals that dTrans occupy similar environments to dIs but have luminosities comparable to dSphs.  We further identify two M81~Group dEs with luminosities, tidal indices, and SFHs which may be analogs to this rare class of galaxy in the LG. The excellent agreement between the two samples, and among the broader universe, underlines the importance of environmental factors in the evolution of dwarf galaxies.  

In summary, we find that the cumulative SFHs and morphology-density relationships of the LG and ANGST samples are not drastically different.  This finding suggests that the LG dwarf galaxies are reasonably representative of dwarf galaxies in the wider universe.    

\acknowledgments

The authors would like to thank the anonymous referee for a number of insightful comments that have helped to improve this paper.  DRW is grateful for support from the University of Minnesota Doctoral Dissertation Fellowship and Penrose Fellowship.  IDK is partially supported by RFBR grant 10-02-00123.  Support for KMG is provided by NASA through Hubble Fellowship grant HST-HF-51273.01 awarded by the Space Telescope Science Institute. This work is based on observations made with the NASA/ESA Hubble Space Telescope, obtained from the data archive at the Space Telescope Science Institute.  Support for this work was provided by NASA through grants GO-10915, DD-11307, and GO-11986 from the Space Telescope Science Institute, which is operated by AURA, Inc., under NASA contract NAS5-26555. This research has made use of the NASA/IPAC Extragalactic Database (NED), which is operated by the Jet Propulsion Laboratory, California Institute of Technology, under contract with the National Aeronautics and Space Administration.  This research has made extensive use of NASA's Astrophysics Data System Bibliographic Services.  
{\it Facility:} \facility{HST (ACS, WFPC2)}

\end{document}